\documentclass[a4paper,12pt]{article}
\usepackage{enumerate}
\usepackage{amsmath,amssymb}
\usepackage[dvips]{graphicx}
\usepackage{color}
\newtheorem{theorem}{Theorem}
\newtheorem{lemma}{Lemma}
\newtheorem{proposition}{Proposition}
\newtheorem{remark}{Remark}
\newcommand{\cB}{\mathcal{B}}
\newcommand{\cF}{\mathcal{F}}
\newcommand{\cH}{\mathcal{H}}
\begin{document}
\title{Scattering through a straight quantum waveguide with combined boundary conditions}
\author{Ph. Briet$^{1,2}$, J. Dittrich$^3$, E. Soccorsi$^{1,2}$\\
{\small $^1$Aix-Marseille Universit\'{e}, CNRS, CPT UMR 7332, 13288 Marseille, France} \\
{\small $^2$Universit\'{e} de Toulon, CNRS, CPT UMR 7332, 83957 La Garde, France} \\
{\small $^3$Nuclear Physics Institute ASCR, CZ-250 68 \v Re\v z, Czech Republic}}
\date{11.8.2014}
\maketitle
\abstract{Scattering through a straight two-dimensional quantum
waveguide $\mathbb{R} \times (0,d)$ with Dirichlet
boundary conditions on $(\mathbb{R}_-^* \times \{ y=0 \} ) \cup (\mathbb{R}_+^* \times \{ y=d \} )$ and
Neumann boundary condition on $(\mathbb{R}_-^* \times \{ y=d \} ) \cup (\mathbb{R}_+^* \times \{ y=0 \} )$ is
considered using stationary scattering theory. The existence of a matching conditions
solution at $x=0$ is proved. The use of stationary scattering theory is justified showing its relation
to the wave packets motion. As an illustration, the matching conditions are also solved numerically and the
transition probabilities are shown. }

\section{Introduction}
Free motion of the non-relativistic quantum mechanical particle is
described by Laplace operator as the Hamiltonian up to unessential
constants, i.e. if units where $\hbar=1$ and particle mass
$m=\frac{1}{2}$ are used. Impenetrable walls guiding the particle
motion are described by boundary conditions on the wave function
annulating the current through the walls and at the same time
making the Laplacian to be a self-adjoint operator with the
suitable domain. Restricting ourselves to the local boundary
conditions, they are of the Robin form with the real coefficient.
The important special cases are Dirichlet and Neumann boundary
conditions known also from other parts of physics. They can also
effectively appear for the wave functions of special symmetry,
e.g. \cite{Evans,Vanek}. If walls with different types of boundary
conditions can be realized e.g. in some semiconductor materials we
would have at our disposal a new type of electron motion control
giving perspectives of new microelectronic elements. The study of
quantum mechanical problems with combined Dirichlet and Neumann
boundary conditions is also a mathematical challenge which can
lead to further solvable or nearly solvable models.

Two-dimensional straight quantum waveguides with the combined Dirichlet and Neumann boundary conditions
are studied for years. They were examined as auxiliary problems in \cite{Evans,Vanek} and the existence of bound states was proved for some configurations. The existence of bound states in a Dirichlet planar waveguide with Neumann window was shown in \cite{Vugalter}.
In \cite{DittKr}, the existence or non-existence of bound states for the cases of Dirichlet and Neumann boundary conditions on half-lines of each boundary were shown in dependence of the overlap of Neumann part projections to the waveguide axis. The Hamiltonian domain was also studied thoroughly. Further details on the discrete spectrum were obtained in \cite{Borisov1}. Paper \cite{Najar} show the existence of bound states for the 3-dimensional layer with Dirichlet boundary conditions outside one or two Neumann disc-shaped windows. The time decay of heat equation solution in two-dimensional waveguides with combined boundary conditions was studied in \cite{Krejcirik}. The cases with infinitely many changes of boundary condition type were studied in \cite{Borisov2}. The limit of infinitely thin waveguide was investigated in \cite{Borisov} and the Dirichlet-like decoupling of parts with different boundary conditions proved in the limit.
In the above mentioned papers, mostly spectral properties were studied.

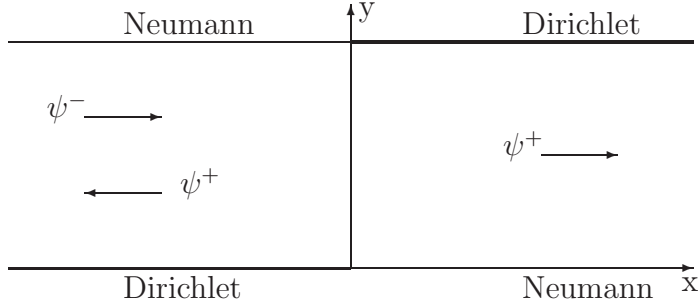
\begin{figure}[h]
\begin{center}
\begin{minipage}[6 cm]{10.5 cm}
\setlength{\unitlength}{0.5 cm}
\begin{picture}(20, 10)
\thicklines
\put(1,1){\line(1,0){9}}
\put(10,7){\line(1,0){9}}
\thinlines
\put(1,7){\line(1,0){9}}
\put(10,1){\vector(1,0){9}}
\put(10,1){\vector(0,1){7}}
\put(4,0.2){Dirichlet}
\put(14.5,7.2){Dirichlet}
\put(4,7.2){Neumann}
\put(14.5,0.2){Neumann}
\put(18.7,0.4){x}
\put(10.2,7.7){y}
\put(15,4){\vector(1,0){2}}
\put(3,5){\vector(1,0){2}}
\put(5,3){\vector(-1,0){2}}
\put(14,4){$\psi^{+}$}
\put(5.5,3){$\psi^{+}$}
\put(2,5){$\psi^{-}$}
\label{wgshape}
\end{picture}
\caption{The shape of waveguide.}
\end{minipage}
\end{center}
\end{figure}

In the present paper we investigate a simple situation of scattering occuring in the planar straight strip-like domain that is displayed on Figure \ref{wgshape}.
The Hamiltonian $H$ is the (minus) Laplace operator in the waveguide $\Omega=\mathbb{R} \times (0,d)$ with
the Dirichlet boundary condition on $( \mathbb{R}_-^*\times\{0\}) \cup (\mathbb{R}_+^*\times \{d\})$ and the Neumann boundary condition on $(\mathbb{R}_-^*\times\{d\}) \cup (\mathbb{R}_+^* \times \{0\})$.
Its domain reads (see \cite{DittKr})
\begin{eqnarray*}
{\mathcal D}(H) & = & \left\{ \psi\in H^1(\Omega) | -\Delta \psi\in L^2(\Omega), \right. \\
& & \left. \psi(-x,0)=\psi(x,d)=\frac{\partial \psi}{\partial y}(x,0)=\frac{\partial \psi}{\partial y}(-x,d)=0\ {\rm for}\ x>0 \right\}.
\end{eqnarray*}
The domain of $H$ is not $H^2(\Omega)$ but it is contained in $H^2_{\rm loc}(\Omega)$. In fact, functions from ${\mathcal D}(H)$ are in $H^2(\Omega_1)$
for any open $\Omega_1\subset \Omega$ such that $\overline{\Omega_1}\cap\{(0,0)\, ,\, (0,d)\}=\emptyset$. This is an analogy of the well known situation for the Dirichlet Laplacian on a planar domains with angles larger than $\pi$ on the boundary \cite{Birman}.

We consider the scattering from the left ($x\to -\infty$) to the right ($x\to +\infty$), the oposite case being symmetric.
As reference (free motion) Hamiltonians for the scattering we use
two (minus) Laplace operators with Dirichlet boundary condition on the whole lower boundary $y=0$ and
Neumann boundary condition on the whole upper boundary $y=d$ or vice versa. Their spectral
decompositions are clear from the variables separation. The corresponding transversal modes are
\begin{equation}
\label{chidef}
\chi^{-}_n(y)=\sqrt{\frac{2}{d}}\sin\left((2n-1)\frac{\pi y}{2d}\right),\
\chi^{+}_n(y)=\sqrt{\frac{2}{d}}\cos\left((2n-1)\frac{\pi y}{2d}\right)
\end{equation}
where $n \in \mathbb{N}^*$, with eigenvalues
$\mu_n=(2n-1)^2\frac{\pi^2}{4d^2}$.
%%%%%%%%%%%%%%%%%%%%%%%%%%%%%%%2%%%%%%%%%%%%%%%%%%%%%%%%%%%%%%%%%%%%%%%%%%%%%%%%%%%%
\section{Stationary scattering wave function}
\label{sec-stationary}
Given an initial transversal mode $n_0 \in \mathbb{N}^*$, we fix the energy $E \in (\mu_N,\mu_{N+1})$ for some natural number $N \geq n_0$ and pick
a longitudinal momentum $k \in \mathbb{R}^*_+$ such that
\begin{equation}
\label{energy}
E=\mu_{n_0}+k^2.
\end{equation}
We skip the case of $E$ equal to the energy of some transversal mode $\mu_n$ for some $n \in \mathbb{N}^*$, whose set has zero Lebesgue measure on the real line generated by $k$.
Let us denote the maximal excitable transversal mode as

\begin{equation}
\label{n1}
n_1(k)=\left[ \frac{1}{2}+\sqrt{\left(n_0-\frac{1}{2}\right)^2+\frac{d^2}{\pi^2}k^2} \right],
\end{equation}
where the symbol $\left[ \cdot \right]$ stands for the integer part, and set
\begin{equation}
\label{kl}
k_n=\sqrt{\mu_{n_0}-\mu_n+k^2} > 0,\ \mbox{for}\ n=1,2,\dots, n_1(k),
\end{equation}
and
\begin{equation}
\label{kn}
\kappa_n=\sqrt{\mu_n-\mu_{n_0}-k^2}>0\ {\rm for}\ n \geq n_1(k)+1.
\end{equation}
In light of \eqref{kn} we may find constants $c_0(k), c_1(k)>0$, depending only on $k$, such that the estimates
\begin{equation}
\label{c_estim}
c_0(k) \leq \frac{\kappa_n(k)}{n} \leq  c_1(k)
\end{equation}
hold for all $n$ sufficiently large.

We note $\cF_{E,n_0,k}$ the set of functions $f(k,\cdot,\cdot)$ of the form
\begin{equation}
\label{fleft}
f(k,x,y)=e^{ikx}\chi^{-}_{n_0}(y) + \sum_{l=1}^{n_1} r_l(k)e^{-ik_l x}\chi^{-}_l(y) + \sum_{n=n_1+1}^\infty r_n(k)e^{\kappa_n x}\chi^{-}_n(y)
\end{equation}
for $x<0$, and
\begin{equation}
\label{fright}
f(k,x,y)=\sum_{l=1}^{n_1}t_l(k)e^{ik_l x}\chi^{+}_l(y) + \sum_{n=n_1+1}^\infty t_n(k)e^{-\kappa_n x}\chi^{+}_n(y)
\phantom{(k)e^{k_n x}\chi^{-}}
\end{equation}
for $x>0$. Here $r_n(k)$ and $t_n(k)$, for $n \in \mathbb{N}^*$, are complex coefficients.

We look for a function $f(k,\cdot,\cdot) \in \cF_{E,n_0,k}$, which is solution to the eigenvalue equation $-\Delta f(k,\cdot,\cdot) = E f(k,\cdot,\cdot)$ in $\Omega$, and satisfies the
following matching conditions
\begin{equation}
\label{matching}
f(k,0^-,\cdot)=f(k,0^+,\cdot),\ \frac{\partial f}{\partial x}(k,0^-,\cdot)=\frac{\partial f}{\partial x}(k,0^+,\cdot).
\end{equation}
We shall justify and precise (\ref{matching}) in subsection \ref{subsec-match}. We require in addition that $f(k,\cdot,\cdot)$ be locally in ${\mathcal D}(H)$, entailing that the restrictions of
$f(k,\cdot,\cdot)$ to $(-L,0)\times (0,d)$ and $(0,L) \times (0,d)$ are respectively in $H^1((-L,0)\times (0,d))$ and $H^1((0,L)\times (0,d))$, for all $L>0$.
Namely, this yields that
\begin{equation}
\label{S1_2}
\sum_{n=1}^\infty n |r_n|^2<\infty\ \mbox{and}\ \sum_{n=1}^\infty n |t_n|^2<\infty.
\end{equation}

%%%%%%%%%%%%%%%%%%%%%%%%%%%%section 2.1%%%%%%%%%%%%%%%%%%%%%%%%%%%%%%%%%%%%%%%%%%%%%%%%
\subsection{Suitable functional spaces}
\label{subsec-fncspaces}
For any $\alpha \in \mathbb{R}$, the space
$$
\cH_{\pm}^{\alpha}=\left\{g=\sum_{n=1}^\infty c_n\chi^{(\pm)}_n\,\; \Big|\; \|g\|_{\alpha,\pm}^2=\sum_{n=1}^\infty n^{2\alpha} |c_n|^2 < \infty \right\}
$$
is Hilbertian for the scalar product $(\cdot,\cdot)_{\alpha,\pm}$ induced by the above defined norm $\| \cdot \|_{\alpha,\pm}$.
For $\alpha\geq 0$, $\cH_{\pm}^{\alpha} \subset L^2(0,d)$ while for $\alpha<0$ this is a space of functionals defined below.
For all $f(k,\cdot,\cdot) \in \cF_{E,n_0,k}$, it can be checked that the trace from left at $x=0$,
\begin{equation}
\label{tleft}
f(k,0^-,y)=\chi^{-}_{n_0}(y) + \sum_{l=1}^{n_1} r_l(k)\chi^{-}_l(y) + \sum_{n=n_1+1}^\infty r_n(k)\chi^{-}_n(y),
\end{equation}
belongs to $\cH_-^{\frac{1}{2}}$, while the trace from right
\begin{equation}
\label{tright}
f(k,0^+,y)=\sum_{n=1}^\infty t_n(k)\chi^{+}_n(y),
\end{equation}
is in $\cH_+^{\frac{1}{2}}$. Identity (\ref{tleft}) (resp., (\ref{tright})) may be verified upon taking finite partial sums in the right hand side of (\ref{fleft}) (resp., (\ref{fright})) as approximating $C^\infty$-functions in $H^1((-L,0)\times (0,d))$ norm (resp., $H^1((0,L)\times (0,d))$ norm) and combining the exponential decay of the corresponding terms for $x \not= 0$ with (\ref{S1_2}).

Imposing on $f(k,\cdot,\cdot)$ that the two expressions in the right hand side of (\ref{tleft}) and (\ref{tright}) coincide in $L^2(0,d)$, the common value $f(k,0^-,\cdot)=f(k,0^+,\cdot)$ then belongs to the space
$\cH^{\frac{1}{2}}=\cH_-^{\frac{1}{2}}\cap \cH_+^{\frac{1}{2}}$.
More generally, for any $\alpha>0$, the space $\cH^{\alpha}=\cH_-^{\alpha}\cap \cH_+^{\alpha}$,
endowed with the scalar product,
$(f,g)_{\alpha}=(f,g)_{\alpha,-} + (f,g)_{\alpha,+}$,
%where, for the sake of notational simplicity, $(f,g)_{\pm}$ stands for $(f,g)_{\frac{1}{2},\pm}$, it is easy to see that $\cH^{\frac{1}{2}}$
is Hilbertian and thus
a reflexive Banach space.
Moreover, the dual space $(\cH_\pm^{\alpha})'$ of $\cH_\pm^{\alpha}$, and $\cH_\pm^{-\alpha}$ are isometric. These two spaces may be identified through the duality bracket
$$ \left( f , g \right)_{\cH^{-\alpha}_\pm,\cH^{\alpha}_\pm}= \sum_{n=1}^{+\infty} \overline{f_n} g_n,\ f=\sum_{n=1}^{+\infty} f_n \chi_n^{\pm},\ g=\sum_{n=1}^{+\infty} g_n \chi_n^{\pm}.$$
Analogously, we put for every $\alpha>0$,
$$
\cH^{-\alpha} = \left( \cH^{\alpha} \right)' \supset \cH^{-\alpha}_- \cup \cH^{-\alpha}_+,
$$
and we write $( f, g )_{-\alpha,\alpha}$ instead of $\left( f , g \right)_{\cH^{-\alpha},\cH^{\alpha}}$.

%%%%%%%%%%%%%%%%%%%%%%%%%%%%%%section 2.2%%%%%%%%%%%%%%%%%%%%%%%%%%%%%%%%%%%%%%%%%%%
\subsection{Matching conditions}
\label{subsec-match}
As we seek for generalized eigenfunctions of the operator $H$, we look for the solution $f$ to the equation $(\Delta +E) f = 0$ in $\Omega$, where $E \in \mathbb R$.

For $k \in \mathbb{R}$ fixed, we know from \eqref{c_estim} that $\kappa_n$ scales like $n$ so the series appearing in \eqref{fleft} (resp., in \eqref{fright})
converges
locally uniformly
in $(-\infty,0) \times (0,d)$ (resp., on $(0,+\infty) \times (0,d)$). Differentiating term by term in \eqref{fleft}-\eqref{fright} we obtain through direct computation that any function $f(k,\cdot,\cdot) \in \cF_{E,n_0,k}$ verifies
$(\Delta + E) f(k,x,y)=0$ for every $(x,y) \in \mathbb{R}_{\pm}^* \times (0,d)$. Moreover,
in the distributional sense in $\mathbb{R}_{\pm}^* \times (0,d)$. For every $\varphi \in C^\infty_0(\Omega)$, it follows from this upon applying Stokes theorem that
the scalar product $( f(k,\cdot,\cdot),(\Delta +E) \varphi )_{L^2(\Omega)}$ expresses as:
\begin{eqnarray}
  \lim_{\varepsilon\downarrow 0} \left(\int_0^d \bar{f}(k,-\varepsilon,y) \frac{\partial \varphi}{\partial x}(-\varepsilon,y)dy
-  \int_0^d \bar{f}(k,\varepsilon,y)\frac{\partial \varphi}{\partial x}(\varepsilon,y)dy
\quad \quad \right. \nonumber\\
- \int_0^d \frac{\partial \bar{f}}{\partial x}(k,-\varepsilon,y) \varphi(-\varepsilon,y)dy
+ \int_0^d \frac{\partial \bar{f}}{\partial x}(k,\varepsilon,y) \varphi(\varepsilon,y)dy .
\quad \label{match0}
\end{eqnarray}
Taking the limit in the above identity and bearing in mind that $f(k,0^\pm,\cdot)\in \cH^{\frac{1}{2}}_{\pm} \subset L^2(0,d)$, we find out that
\begin{eqnarray}
& & ( f(k,\cdot,\cdot) , (\Delta +E) \varphi )_{L^2(\Omega)} \nonumber \\
& & \quad =  \left( f(k,0^-,\cdot)-f(k,0^+,\cdot),\frac{\partial \varphi}{\partial x}(0,\cdot) \right)_{L^2(0,d)} \nonumber \\
& & \quad \quad + ( f_x(k,0^+,\cdot) - f_x(k,0^-,\cdot),\varphi(0,\cdot))_{-\frac{1}{2},\frac{1}{2}},
\label{match1}
\end{eqnarray}
for all $\varphi \in C^\infty_0(\Omega)$. Here $\varphi(0,\cdot)$ is regarded as an element of $\cH^{\frac{1}{2}}$, and
\begin{eqnarray}
f_x(k,0^-,\cdot)&=& ik\chi^{-}_{n_0} -i \sum_{l=1}^{n_1} k_l r_l(k)\chi^{-}_l
+ \sum_{n=n_1+1}^\infty \kappa_n r_n(k)\chi^{-}_n  \, \in \, \cH^{-\frac{1}{2}}_-,
\nonumber \\
\label{F-trace}
\\
f_x(k,0^+,\cdot)&=&
i \sum_{l=1}^{n_1}k_l t_l(k)\chi^{+}_l- \sum_{n=n_1+1}^\infty \kappa_n t_n(k)\chi^{+}_n \, \in \, \cH^{-\frac{1}{2}}_+.
\label{F+trace}
\end{eqnarray}
The two above expressions are obtained upon formally differentiating at $x=0$ each term in the series (\ref{fleft})-(\ref{fright}) with respect to $x$. Actually, (\ref{F-trace})-(\ref{F+trace}) may be rigorously justified with the help of (\ref{kn}) and (\ref{S1_2}). We refer to subsection \ref{sec-scndmatching} in Appendix for more details. In light of \eqref{match1} we have obtained the following result.
\begin{proposition}
\label{matchtheor}
For $n_0 \in \mathbb{N}^*$ fixed, pick $E \in (\mu_{n_0},\infty)$ and $k \in \mathbb{R^*_+}$ in accordance with (\ref{energy}) and let
$f(k,\cdot,\cdot) \in \cF_{E,n_0,k}$. Assume moreover that the two following conditions holds simultaneously:
\begin{eqnarray}
& f(k,0^-,\cdot)=f(k,0^+,\cdot) & \mbox{in}\ \cH^{\frac{1}{2}} \label{fcontinuity}\\
& f_x(k,0^-,\cdot)=f_x(k,0^+,\cdot) & \mbox{in}\ \cH^{-\frac{1}{2}} \label{dfcontinuity}
\end{eqnarray}
where $f_x(k,0^\pm,\cdot)$ is defined in (\ref{F-trace})-(\ref{F+trace}).
Then we have $(\Delta +E) f(k,\cdot,\cdot)=0$ in the distributional sense in $\Omega$.
\end{proposition}

\begin{remark}  The equality (\ref{fcontinuity}) is a necessary condition for the conclusion of Proposition \ref{matchtheor} to hold, but we do not know whether this is the case for
(\ref{dfcontinuity}). This can be seen upon taking
$\varphi(x,y)=\varphi_1(x) \varphi_2(y)$ in (\ref{match1}) for two arbitrary functions $\varphi_1\in C_0^\infty({\mathbb R})$ and $\varphi_2\in C_0^\infty(0,d)$, getting
that both terms in the right hand side of (\ref{match1}) vanish. This entails that $f(k,0^-,\cdot)=f(k,0^+,\cdot)$ in $L^2(0,d)$, which yields (\ref{fcontinuity}), and that
the restrictions of $f_x(k,0^-,\cdot)$ and $f_x(k,0^+,\cdot)$ to $C_0^\infty(0,d)$ coincide. As we did not succeed to prove that
$C_0^\infty(0,d)$ is dense in $\cH^{\frac{1}{2}}$, it is not clear whether (\ref{dfcontinuity}) can be derived from this.
\end{remark}

%%%%%%%%%%%%%%%%%%%%%%%%%%%%%%section2.3%%%%%%%%%%%%%%%%%%%%%%%%
%
\subsection{Existence result}
This section is devoted to proving the existence of an actual $f(k,\cdot,\cdot) \in \cF_{E,n_0,k}$ satisfying the stationary Schr\"odinger equation $(\Delta +E) f =0$ in the distributional sense in $\Omega$.
In light of Proposition \ref{matchtheor}, it is enough to seek for a function $f(k,\cdot,\cdot)$ in $\cF_{E,n_0,k}$ fulfilling the matching conditions (\ref{fcontinuity})-(\ref{dfcontinuity}).  With reference to (\ref{fcontinuity}), we denote by $\varphi=f(k,0^-,\cdot)=f(k,0^+,\cdot)$ the common value in $\cH^{\frac{1}{2}}$ of traces  from left and right at $x=0$
of $f(k,\cdot,\cdot)$. In view of (\ref{fleft})-(\ref{fright}), we have
\begin{equation}
\label{phiexpansion}
\varphi=\chi^{-}_{n_0} + \sum_{n=1}^\infty r_n(k) \chi^{-}_n
=\sum_{n=1}^\infty t_n(k) \chi^{+}_n \in \cH^{\frac{1}{2}}.
\end{equation}
Relation (\ref{phiexpansion}) holds necessarily in $L^2(0,d)$, expressing the function $\varphi$ in two different Hilbertian basis of this space, so there is one-to-one
correspondence between the coefficients $r_n(k)$ and $t_n(k)$. We are thus left with the task of finding  a function $\varphi$ of the form prescribed by (\ref{phiexpansion}) which fulfills (\ref{dfcontinuity}).
To this purpose we put for all $n \in \mathbb{N}^*$,
\begin{equation} \label{projectors}
p_n^\pm=(\chi^{\pm}_n,\cdot)_{L^2(0,d)} \chi^{\pm}_n,
\end{equation}
and  we recall (\ref{F-trace})-(\ref{F+trace}) and (\ref{phiexpansion}) in order to re-express $f_x(k,0^\pm,\cdot)$ in terms of $\varphi$, getting:
\begin{eqnarray*}
 f_x(k,0^-,\cdot) &=& \left( -i\sum_{l=1}^{n_1}k_l p_l ^-+ \sum_{n=n_1+1}^\infty \kappa_n p_n^- \right) \varphi + 2ik \chi^{-}_{n_0}, \\
f_x(k,0^+,\cdot) &=& \left( i\sum_{l=1}^{n_1} k_l p_l^+ - \sum_{n=n_1+1}^\infty \kappa_n p_n^+ \right) \varphi.
\end{eqnarray*}
In light of the two above identities, the matching condition (\ref{dfcontinuity}) may be equivalently reformulated as
\begin{equation}
\label{basicmatch}
D\varphi = - 2ik \chi^{-}_{n_0},
\end{equation}
where $D : \cH^{\frac{1}{2}} \to \cH^{-\frac{1}{2}}$ is the operator
\begin{equation}
\label{Ddef}
D=-iD_1+D_2,\
D_1 = \sum_{l=1}^{n_1} k_l (p_l^- + p_l^+),\ D_2=\sum_{n=n_1+1}^\infty \kappa_n (p_n^- + p_n^+).
\end{equation}
Since $k_n \slash n$ and $\kappa_n \slash n$ are both majorized by $2 \sqrt E$, uniformly in $n \in \mathbb{N}^*$, then the estimate
$$ \|D\psi\|_{-\frac{1}{2}}  \leq 2 \sqrt E \|\psi\|_{\frac{1}{2}} $$
holds true for any  $\psi = \sum_{n \geq 1} a_n \chi_n^{-} = \sum_{n \geq 1} b_n \chi_n^{+} \in \cH^{\frac{1}{2}}$. As a consequence, the linear operator $D$ is bounded from $\cH^{\frac{1}{2}}$ into
$\cH^{-\frac{1}{2}}$, with $\|D\|_{\cB(\cH^{\frac{1}{2}},\cH^{-\frac{1}{2}})} \leq 2 \sqrt E$.

The proof of the existence of $\varphi$ obeying (\ref{basicmatch}) boils down to the following technical but essential result, whose proof is postponed to subsection \ref{sec-prlmD2} in Appendix.

\begin{lemma}
\label{lm-D2}
The operator $D_2$ is a strictly positive and boundedly invertible from $\cH^{\frac{1}{2}}$ onto $\cH^{-\frac{1}{2}}$.
Moreover, the restriction $(D_2^{-1})_{| L^2(0,d)}$ of $D_2^{-1}$ to $L^2(0,d)$, is symmetric.
\end{lemma}

In light of (\ref{Ddef}), the matching condition (\ref{basicmatch}) reads
$-iD_1 \varphi + D_2 \varphi = - 2ik \chi^{-}_{n_0}$ so it may be equivalently rephrased as
\begin{equation}
\label{invD2eq}
\varphi -i D_2^{-1} D_1 \varphi = - 2ik D_2^{-1} \chi^{-}_{n_0},
\end{equation}
according to Lemma \ref{lm-D2}.

For every $n \in \mathbb{N}^*$, we denote by $P_n$ the orthogonal projection onto the finite dimensional subspace
${\mathcal P}_n = {\rm span}\{\chi^{\pm}_j,\ j=1,\ldots, n \}$.
Due to (\ref{kl}), $D_1$ is a positive self-adjoint bounded operator in ${\mathcal P}_{n_1}$ and the same is true for $D_1^{1 \slash 2}$ and its inverse $D_1^{-1 \slash 2}$. Further, putting
$$ \varphi_1 = D_1^{1 \slash 2} P_{n_1} \varphi, $$
%in such a way that $\varphi_j \in \cH^{\frac{1}{2},}_0 \subset L^2(0,d)$ for $j=1,2$, and bearing in mind that $D_1 %\varphi_2=0$ so that
we may rewrite (\ref{invD2eq}) as
$\varphi - i D_2^{-1} D_1^{1 \slash 2} \varphi_1 = - 2ik D_2^{-1} \chi^{-}_{n_0}$.
Applying $D_1^{1 \slash 2} P_{n_1}$ and $I-P_{n_1}$ successively to both sides of the above equation, we end up getting that (\ref{invD2eq}) is equivalent to the system formed by the two following equations:
\begin{eqnarray}
(I - i D_1^{1 \slash 2} P_{n_1} D_2^{-1} P_{n_1} D_1^{1 \slash 2} ) \varphi_1 &= & -2ik D_1^{1 \slash 2} P_{n_1} D_2^{-1}\chi^{-}_{n_0} \label{phi1eq} \\
\varphi_2 - i (I-P_{n_1}) D_2^{-1} P_{n_1} D_1^{1 \slash 2} \varphi_1 & = & -2ik (I-P_{n_1}) D_2^{-1} \chi^{-}_{n_0}, \label{phi2eq}
\end{eqnarray}
where
$$ \varphi_2 = (I-P_{n_1}) \varphi. $$
Notice from the second part of Lemma \ref{lm-D2} that the operator $M=D_1^{1 \slash 2} P_{n_1} D_2^{-1} P_{n_1} D_1^{1 \slash 2}$ is symmetric in ${\mathcal P}_{n_1}$.
%As $D_2(\bar\varphi)(\varphi) \geq \eta_{n_1}^2 \|\varphi\|_{L^2}$, positivity of $M$ is easily verified. So $M$ acts as
%a positive Hermitian matrix in ${\mathcal P}_{n_1}$ and
As a consequence, all the eigenvalues $\lambda_j$, $j=1,\dots,n_1$, of $M$, are real and $\det(I-iM)=\prod_{j=1}^{n_1} (1-i\lambda_j) \not= 0$, showing that
(\ref{phi1eq}) admits a unique solution $\varphi_1 \in {\mathcal P}_{n_1} \subset L^2(0,d)$. From this and (\ref{phi2eq}) then follows that $\varphi_2 \in L^2(0,d)$, which in turn yields
$$
\varphi = D_1^{-\frac{1}{2}}\varphi_1 + \varphi_2 = i D_2^{-1}(P_{n_1} D_1^{1 \slash 2} \varphi_1 - 2k\chi^{-}_{n_0}) \in \cH^{\frac{1}{2}},
$$
since $L^2(0,d) \subset \cH^{-\frac{1}{2}}$. This proves the existence of $\varphi$ and consequently of $f$. Summing up, we have obtained the:
\begin{theorem}
\label{f_existence}
Let $n_0$, $E$ and $k$ be the same as in Proposition \ref{matchtheor}. Then there exists $f(k,\cdot,\cdot) \in \cF_{E,n_0,k}$ satisfying
the equation $(\Delta +E)f(k,\cdot,\cdot)=0$ in the distributional sense in $\Omega$.
\end{theorem}

\begin{remark} Let $\psi \in \cH^{-\frac{1}{2}}$ be arbitrary. It is clear that the above reasonning remains valid upon substituting $\psi$ for $-2 i k \chi_{n_0}^-$ in the right hand side of \eqref{basicmatch}. Therefore the equation $D \varphi = \psi$ admits a unique solution
solution $\varphi \in\cH^{\frac{1}{2}}$. This entails that the linear bounded operator $D$ is invertible from $\cH^{\frac{1}{2}}$ onto $\cH^{-\frac{1}{2}}$. By the
%open
inverse mapping theorem, its inverse $D^{-1}$ is thus bounded from $ \cH^{-\frac{1}{2}}$ onto $\cH^{\frac{1}{2}}$.
\end{remark}

%%%%%%%%%%%%%%%%%%%%%%%%%%%%%%section 2.5%%%%%%%%%%%%%%%%%%%%%%%%%%
\subsection{Smooth $k$-dependence}
For all $(x,y) \in \mathbb{R}^* \times (0,d)$, it is apparent from \eqref{fleft}-\eqref{fright} that $k \mapsto f(k,x,y)$ is continuously differentiable in $\mathbb{R} \setminus M$, where the set
$$
M=\{ k\in \mathbb{R}_+^*\,|\, \exists n \in \mathbb{N}^*,\ \mu_n = k^2+\mu_{n_0} \}
$$
is discrete. Moreover, for $x=0$ we have the:
\begin{theorem}
\label{dphidk}
The solution to (\ref{basicmatch}), regarded as a function of the parameter $k \in \mathbb{R}_+^*$, belongs to $C^1(\mathbb{R}_+^* \setminus M ; \cH^{\frac{1}{2}})$.
\end{theorem}
{ \bf Proof.} We prove that $k \mapsto f(k,0,y)$ is continuously differentiable for any $y \in (0,d)$.
To this purpose we first estimate the variation of the operators $D=D(k)$ with respect to $k$. To do that we fix $k \in \mathbb{R}$ and refer to \eqref{kl}-\eqref{kn} in order to
choose $\varepsilon_1=\varepsilon_1(k) >0$ so small that $n_1(k+\varepsilon)=n_1(k)$ for $\varepsilon \in (-\varepsilon_1 ,\varepsilon_1)$. Then we apply \eqref{Ddef}, getting
\begin{eqnarray}
D(k+\varepsilon)-D(k)
& = & -i\sum_{l=1}^{n_1}(k_l(k+\varepsilon)-k_l(k)) (p_l^- + p_l^+) \nonumber \\
& & + \sum_{n=n_1+1}^\infty (\kappa_n(k+\varepsilon)-\kappa_n(k)) (p_n^-+p_n^+), \quad \label{Depsilon}
\end{eqnarray}
where $n_1$ stands for $n_1(k)=n_1(k+\varepsilon)$.
Next we recall from \eqref{kl}-(\ref{kn}) that there exist a continuous functions $c_n(k,\cdot)$ obeying
\begin{eqnarray*}
k_n(k+\varepsilon)-k_n(k) & = & \frac{k}{k_n(k)}\varepsilon + c_n(k,\varepsilon)\varepsilon^2,\
n=1,.., n_1, \\
\kappa_n(k+\varepsilon)-\kappa_n(k) & = & -\frac{k}{\kappa_n(k)}\varepsilon + c_n(k,\varepsilon)\varepsilon^2,\ n \geq n_1 + 1,
\end{eqnarray*}
for $(- \varepsilon_1 ,\varepsilon_1)$. Moreover, we know from (\ref{c_estim}) that the estimate
$|c_n(k,\varepsilon)|\leq c(k)  / n$ holds uniformly in $\varepsilon \in (-\varepsilon_1,\varepsilon_1)$, where the constant $c(k)>0$ depends only on $k$ and $\varepsilon_1$.
Hence
$$
C(k,\varepsilon)=\sum_{n=1}^\infty c_n(k,\varepsilon)(p_n^-+p_n^+) \in \mathcal{B}(\cH^{\frac{1}{2}},\cH^{-\frac{1}{2}}),\
$$
with
\begin{equation}
\label{quadrep}
\|C(k,\varepsilon)\|_{\mathcal{B}(\cH^{\frac{1}{2}},\cH^{-\frac{1}{2}})} \leq c(k),\ \varepsilon \in (-\varepsilon_1,\varepsilon_1).
\end{equation}
 for some positive constant $c(k)$ independent of $\varepsilon$. Arguing in the same way as above we obtain that the linear term with respect to $\varepsilon$, appearing in (\ref{Depsilon}), satisfies
\begin{equation}
\label{linep}
\left\|-ik\sum_{l=1}^{n_1}\frac{1}{k_l}(p_l^-+p_l^+) - k \sum_{n=n_1+1}^\infty\frac{1}{\kappa_n}(p_n^-+p_n^+)\right\|_{\mathcal{B}(\cH^{\frac{1}{2}},\cH^{-\frac{1}{2}})} \leq c(k),
\end{equation}
where $c(k)$ is another constant depending only on $k$. Putting (\ref{Depsilon}) and (\ref{quadrep})-(\ref{linep}) together we obtain for all $|\varepsilon| < \varepsilon_1$ that
$$ \|D(k+\varepsilon) - D(k)\|_{\mathcal{B}(\cH^{\frac{1}{2}},\cH^{-\frac{1}{2}})} \leq c(k) \varepsilon (1 + \varepsilon), $$
hence $\|D(k)^{-1}(D(k+\varepsilon) - D(k))\|_{\mathcal{B}(\cH^{\frac{1}{2}})} <1$, upon eventually shortening $\varepsilon_1$. Therefore $I+D(k)^{-1}(D(k+\varepsilon) - D(k))$ is invertible in $\mathcal{B}(\cH^{\frac{1}{2}})$ for every $\varepsilon \in (-\varepsilon_1,\varepsilon_1)$ and the same is true for
$D(k+\varepsilon)=D(k) (I+D(k)^{-1}(D(k+\varepsilon) - D(k)))$ in ${\mathcal{B}(\cH^{\frac{1}{2}},\cH^{-\frac{1}{2}})}$. Moreover $\Vert D(k+\varepsilon)^{-1} \Vert_{\mathcal{B}( \cH^{-\frac{1}{2}},\cH^{\frac{1}{2}})}$ is uniformly bounded by, say, $c'= c'(k)>0$, in $\varepsilon \in (-\varepsilon_1,\varepsilon_1)$. Thus, bearing in mind that
\begin{eqnarray*}
& & \|D(k+\varepsilon)^{-1}-D(k)^{-1}\|_{\mathcal{B} ( \cH^{-\frac{1}{2}},\cH^{\frac{1}{2}} )} \\
& = & \|D(k)^{-1} (D(k)-D(k+\varepsilon)) D(k+\varepsilon)^{-1}\|_{\mathcal{B} ( \cH^{-\frac{1}{2}},\cH^{\frac{1}{2}} )} \\
& \leq & c'^2 \| (D(k)-D(k+\varepsilon)) \|_{\mathcal{B} ( \cH^{\frac{1}{2}},\cH^{-\frac{1}{2}} )},
\end{eqnarray*}
we end up getting that
\begin{equation}
\label{Dinvcont}
\lim_{\varepsilon \to 0} \| D(k+\varepsilon)^{-1}-D(k)^{-1} \|_{\mathcal{B} ( \cH^{-\frac{1}{2}},\cH^{\frac{1}{2}} )} =  0.
\end{equation}
For $\varepsilon \in (-\varepsilon_1, \varepsilon_1)$, let $\varphi(k+\varepsilon)=D(k+\varepsilon)^{-1}(-2 i (k+\varepsilon) \chi_{n_0}^{-})$ denotes the solution to (\ref{basicmatch}).
Then we have $D(k+\varepsilon) ( \varphi(k+\varepsilon)-\varphi(k) ) = - (D(k+\varepsilon)-D(k))\varphi(k) - 2i\varepsilon\chi^{-}_{n_0}$ by direct calculation, hence
\begin{eqnarray*}
& & \frac{\varphi(k+\varepsilon)-\varphi(k)}{\varepsilon} \\
& = & - D(k+\varepsilon)^{-1} \partial_k D(k) \varphi(k) - 2iD (k+\varepsilon)^{-1}\chi^{-}_{n_0}
-\varepsilon D(k+\varepsilon)^{-1} C(k,\varepsilon) \varphi(k),
\end{eqnarray*}
for any $\ 0 < | \varepsilon | < \varepsilon_1$, where
$$ \partial_k D(k) = -ik \sum_{l=1}^{n_1}\frac{1}{k_l}(p_l^-+p_l^+) - k \sum_{n=n_1+1}^\infty\frac{1}{\kappa_n}(p_n^-+p_n^+). $$
From this, (\ref{quadrep}) and (\ref{Dinvcont}) then follows that $\varphi$ is differentiable in $\cH^{\frac{1}{2}}$ at $k$, with
$$
\frac{\partial \varphi}{\partial k} (k) = -D(k)^{-1} \partial_k D(k) \varphi(k) - 2iD(k)^{-1}\chi^{-}_{n_0} \in \cH^{\frac{1}{2}}.
$$
Notice that this expression coincides with the one obtained by formal differentiation of (\ref{basicmatch}).
Its continuity with respect to $k$ was actually established in the proof.
\begin{flushright} $\square$ \end{flushright}
\begin{remark}
By induction, the continuity for higher derivatives of $k \mapsto \varphi(k)$ follows from the reasoning developped in the proof of Theorem \ref{dphidk}.
\end{remark}

%%%%%%%%%%%%%%%%%%%%%%%%%%%%%%%%%%%%%%%%%%%%%%%%%%%%%%%%%%%%%%%%%%%%%%%%%%%%%%%%%%%%%%%%%%%%%%%%%%%%%%%%%%%%%%%%%%%%%%%%%%%%%%%%%%%%%%%%%%%%%
\section{Time-dependent solutions}
We now characterize the scattering properties of the physical model under study with the aid of the generalized eigenfunctions of $H$, described in section \ref{sec-stationary} (see Theorem \ref{f_existence}). For the sake of simplicity we consider sufficiently small quasi-momenta intervals $[\alpha,\beta]$, with $0 < \alpha < \beta$, so that the index $n_1(k)$ of the highest excited transversal mode remains constant with respect to $k \in [\alpha, \beta]$. For any fixed $k_0 \in \mathbb{R}$, it is clear from (\ref{n1}) that this can be achieved upon imposing
\begin{equation}
\label{time0a}
\sqrt{\mu_{n_1(k_0)} -\mu_{n_0}} < \alpha < \beta < \sqrt{\mu_{n_1(k_0)+1} -\mu_{n_0}}.
\end{equation}
Let $a\in C^\infty_0({\mathbb R})$ be such that
\begin{equation}
\label{time0b}
{\rm supp}\ a \subset [\alpha,\beta] \subset (0,\infty)\setminus M,
\end{equation}
where $M$ is as in Theorem \ref{dphidk}, and let $f \in \mathcal{F}_{E,n_0,k}$ be defined by Theorem \ref{f_existence}. In view of characterizing the asymptotic behaviour of the time-evolution generated by $H$ of
\begin{equation}
\label{time1}
\psi(t,x,y)=\int_{\mathbb R} a(k) e^{-i(\mu_{n_0}+k^2)t}f(k,x,y) dk,
\end{equation}
we introduce the two following states,
\begin{equation}
\label{time2}
\psi^{-}(t,x,y)=\int_{\mathbb R}a(k)e^{-i(\mu_{n_0}+k^2)t}e^{ikx}\chi^{-}_{n_0}(y)\,dk
\end{equation}
and
\begin{equation}
\label{time3}
\psi^{+}(t,x,y)= \left\{ \begin{array}{cl} \psi_{DN}(t,x,y) & \mbox{if}\ x<0, \\ \psi_{ND}(t,x,y) & \mbox{if}\ x>0, \end{array}
\right.
\end{equation}
where we have set
\begin{eqnarray}
\psi_{DN}(t,x,y) & = & \int_{\mathbb R}a(k)e^{-i(\mu_{n_0}+k^2)t} \sum_{l=1}^{n_1} r_l(k)e^{-ik_l x}\chi^{-}_l(y) dk, \label{time4} \\
\psi_{ND}(t,x,y) & = & \int_{\mathbb R}a(k)e^{-i(\mu_{n_0}+k^2)t} \sum_{l=1}^{n_1} t_l(k)e^{ik_l x}\chi^{+}_l(y) dk. \label{time5}
\end{eqnarray}
As can be seen from the following statement,
$\psi_{DN}$ (resp. $\psi_{ND}$) is a weak solution to the Schr\"odinger equation associated with the Laplace operator $\Delta_{DN}$ (resp. $\Delta_{ND}$) acting in $L^2(\Omega)$ with the Dirichlet (resp. Neumann) boundary conditions at $y=0$ and Neumann (resp. Dirichlet) boundary conditions at $y=d$.

%%%%%%%%%%%%%%%%%%%%%%%%%%%%%%%%%%%%%%%%%%%%%%%%%%%%%%%%%%%%%
\begin{proposition}
\label{pr-2}
For every $t \in \mathbb{R}$, the function $\psi(t,\cdot,\cdot)$, defined by (\ref{time1}), belongs to
${\mathcal D}(H)$.
% $$ {\mathcal D}_{\rm loc}(H):= \{ u \in L_{\rm{loc}}^2(\Omega),\ \forall \eta \in % C_0^{\infty}(\mathbb{R}),\ \eta(x) u(x,y) \in {\mathcal D}(H) \}. $$
Moreover $\psi$ is a solution
% in the distributional sense
to the equation
\begin{equation}
\label{psi}
\left( i\frac{\partial}{\partial t} + H \right) \psi = 0 .
\end{equation}
Similarly $\psi^-(t,\cdot,\cdot) \in {\mathcal D}(\Delta_{DN})$ for each $t \in \mathbb{R}$ and we have
\begin{equation}
\label{psi-}
\left( i\frac{\partial}{\partial t} + \Delta_{DN}\right) \psi^{-} = 0 .
%\ \mbox{in}\ (C_0^{\infty}(\mathbb{R} \times \Omega))'.
\end{equation}
Finally $\psi_{DN}(t,\cdot,\cdot)$ and $\psi_{ND}(t,\cdot,\cdot)$ are respectively in ${\mathcal D}(\Delta_{DN})$ and in ${\mathcal D}(\Delta_{ND})$ for all $t \in \mathbb{R}$, with
\begin{equation}
\label{psi+}
\left( i\frac{\partial}{\partial t} + \Delta_{DN}\right) \psi_{DN} =
\left( i\frac{\partial}{\partial t} + \Delta_{ND}\right) \psi_{ND} =0 .
\end{equation}
\end{proposition}
{\bf Proof.}
We only show the first part of the statement, the remaining part being obtained by arguing in the same way.

From the expression (\ref{phiexpansion}) of $\varphi=f(k,0^-,\cdot)=f(k,0^+,\cdot)$ we get that $r_n(k) = ( \chi^{-}_n(\cdot),\varphi (k,\cdot))_{L^2(0,d)}$ (up to some additive constant for $n= n_0$) and
 $t_n(k) = (\chi^{+}_n(\cdot),\varphi (k,\cdot))_{L^2(0,d)}$ for any $n \in \mathbb{N}^*$. Each $r_n$ and $t_n$ as well as $\|\varphi\|_\frac{1}{2}$ is thus a continuous function of $k \in [\alpha, \beta]$ by Theorem \ref{dphidk}. Therefore there exists a constant $c_1>0$ such that we have
\begin{equation}
\label{coeff_estim}
|r_n(k)|\leq c_1 n^{-\frac{1}{2}}\quad \mbox{and}\quad |t_n(k)|\leq c_1 n^{-\frac{1}{2}},
\end{equation}
for all $n \geq 1$ and $k \in [\alpha,\beta]$, by \eqref{S1_2}. On the other hand \eqref{kn} yields
$\kappa_n \geq c_2 n$ uniformly in $n \geq n_1+1$ and $k \in [\alpha, \beta]$, for another positive constant $c_2$.
As a consequence the function defined by
\begin{equation}
\label{psi2}
\tilde{\psi}_{ND}(t,x,y)= \int_{\mathbb R}a(k)e^{-i(\mu_{n_0}+k^2)t} \sum_{n=n_1+1}^\infty t_n(k) e^{-\kappa_n x} \chi^{+}_n(y)  dk,
\end{equation}
for every $(t,x,y) \in \mathbb{R} \times \mathbb{R}_+^* \times (0,d)$, satisfies
\begin{eqnarray*}
\int_0^d | \tilde{\psi}_{ND}(t,x,y)|^2 dy & = & \sum_{n=n_1+1}^\infty \left| \int_{\mathbb R}a(k)e^{-i(\mu_{n_0}+k^2)t}t_n(k) e^{-\kappa_n x} \, dk \right|^2 \\
& \leq & c_1^2 \|a\|_{L^1(\mathbb{R})}^2 \sum_{n=n_1+1}^\infty \frac{e^{-2 c_2 n  x}}{n},\ (t,x) \in \mathbb{R} \times \mathbb{R}_+^*,
\end{eqnarray*}
whence
$$
\int_0^\infty \int_0^d | \tilde{\psi}_{ND}(t,x,y)|^2 dy dx \leq \frac{c_1^2 \|a\|_{L^1(\mathbb{R})}^2}{2 c_2}  \left( \sum_{n=n_1+1}^\infty \frac{1}{n^2} \right)< \infty.
$$
Therefore $\tilde{\psi}_{ND}(t,\cdot,\cdot) \in L^2(\mathbb{R}_+^* \times (0,d))$ for any $t \in \mathbb{R}$, and we get in the same way that $\tilde{\psi}_{DN}(t,\cdot,\cdot) \in L^2( \mathbb{R}_-^* \times (0,d))$, where we have set
\begin{equation}
\label{psi1}
\tilde{\psi}_{DN}(t,x,y)= \int_{\mathbb R} a(k)e^{-i(\mu_{n_0}+k^2)t} \sum_{n=n_1+1}^\infty r_n(k) e^{\kappa_n x} \chi^{-}_n(y) dk,
\end{equation}
for all $(t,x,y) \in \mathbb{R} \times \mathbb{R}_-^* \times (0,d)$.
By performing the change of integration variable $k \to k_l$ in each term of the sum appearing in (\ref{time4})-(\ref{time5}), we obtain the Fourier transform of an $L^2$-function (with respect to the variable $k_l$), which is consequently square integrable with respect to $x$. Therefore the functions defined in
(\ref{time4})-(\ref{time5}) are lying in $L^2(\Omega)$, and we have $\psi \in L^2(\Omega)$ from \eqref{fleft}-\eqref{fright} and \eqref{time1}.

The next step of the proof is to check out that the first order partial derivatives of $\psi(t,\cdot,\cdot)$ are square integrable in $\Omega$ for any $t \in \mathbb{R}$. We shall do it for $\partial \psi \slash \partial y$, the case of $\partial \psi \slash \partial x$ being handled in the same manner. We start by computing $\partial \psi \slash \partial y$ in the
distributional sense. We have
\begin{eqnarray}
& & \langle \frac{\partial \psi}{\partial y}(t,\cdot ,\cdot ) , \zeta \rangle_{(C_0^{\infty}(\Omega))',C_0^{\infty}(\Omega)} \nonumber \\
& = & -\int_\Omega \left(\int_{\mathbb{R}} a(k)e^{-i(\mu_{n_0}+k^2)t} f(k,x,y) dk \right) \frac{\partial \zeta}{\partial y}(x,y)dx dy,
\label{Ldistrib}
\end{eqnarray}
for any arbitrary test function $\zeta \in C_0^\infty (\Omega)$.
For $y \in (0,d)$ and $k \in \mathbb{R}$ fixed, we deduce from \eqref{fright} that $|f(k,x,y)|$ is majorized by
\begin{eqnarray*}
& & \left( \frac{2}{d} \right)^{1 \slash 2} \left( \sum_{l=1}^{n_1}|t_l(k)| + \sum_{n=n_1+1}^\infty |t_n(k)| e^{-c_2 n x} \right) \\
& \leq & \left( \frac{2}{d} \right)^{1 \slash 2} \left( \sum_{l=1}^{n_1}|t_l(k)| +  (1-e^{-2 c_2 x})^{-\frac{1}{2}} \left( \sum_{n=n_1+1}^\infty |t_n(k)|^2 \right)^{1 \slash 2} \right),
\end{eqnarray*}
for all $x>0$, and a similar estimate holds from \eqref{fleft} for $x<0$. Hence the integral over
$({\rm supp}\ \zeta) \times ({\rm supp}\ a)$ in the right hand side of (\ref{Ldistrib}) converges so we may apply Fubini's theorem. By integrating by parts over $y$, we find out by direct calculation that
$$
\frac{\partial \psi}{\partial y}(t ,x ,y)=\int_{\mathbb R} a(k)e^{-i(\mu_{n_0}+k^2)t}f_y(k,x,y) dk,
$$
in the distributional sense, where $f_y(k,x,y)$ expresses as
$$\mu_{n_0}^{1/2}e^{ikx}\chi^{+}_{n_0}(y) + \sum_{l=1}^{n_1}  \mu_{l}^{1/2}r_l(k)e^{-ik_l x}\chi^{+}_l (y)
+ \sum_{n=n_1+1}^\infty \mu_{n}^{1/2} r_n(k) e^{\kappa_n x}\chi^{+}_n(y), $$
for $x<0$, and as
$$- \left( \sum_{l=1}^{n_1} \mu_{l}^{1/2}t_l(k)e^{ik_l x}\chi^{-}_l(y) + \sum_{n=n_1+1}^\infty  \mu_{n}^{1/2}t_n(k)e^{-\kappa_n x}\chi^{-}_n(y) \right),$$
for $x>0$.
%Due to just mentioned form of  $\psi$ we consider  separately $ \partial \psi_2/\partial y$ and $ \partial \psi_1/\partial y$. For $ x>0$ and $ y \in (0,d)$ its holds pointwise
We are thus left with the task of checking out that the two above expressions lead to square integrable
functions in $\Omega$.
Since this is obviously the case for all terms expressed by a finite sum, we shall only examine the one of
$$
\hat{\psi}_{ND}(t,x,y) = \sum_{n=n_1+1}^\infty\mu_{n}^{1/2} \chi^{-}_n(y)
 \int_{\mathbb R}a(k)e^{-i(\mu_{n_0}+k^2)t} t_n(k) e^{-\kappa_n x} dk .
$$
Actually, it turns out that
$$
\int_0^d \left|\hat{\psi}_{ND}(t,x,y)  \right|^2  dy= \sum_{n=n_1+1}^\infty \mu_n
\left| \int_{\mathbb R}a(k)e^{-i(\mu_{n_0}+k^2)t} t_n(k) e^{-\kappa_n x}\, dk \right|^2
$$
so we have
\begin{eqnarray*}
& & \int_0^\infty \int_0^d \left| \hat{\psi}_{ND}(t,x,y)  \right|^2  dxdy \\
& \leq &
\|a\|_{L^2(\mathbb{R})}^2 \int_{{\rm supp}\, a}\sum_{n=n_1+1}^\infty \frac{\mu_n}{2\kappa_n(k)}|t_n(k)|^2 dk < \infty,
\end{eqnarray*}
the last series being majorized (up to some multiplicative constant) by $\|\varphi(k)\|_\frac{1}{2}^2$, which is bounded on
${\rm supp}\, a$.
Now, bearing in mind that $f(k,\cdot,\cdot)$ is locally in the domain of the Laplacian (this is guaranteed by the matching conditions (\ref{fcontinuity})-(\ref{dfcontinuity})), so that $x \mapsto f(k,\cdot,y) \in C^1(\mathbb{R})$ for almost every $y\in (0,d)$ and any arbitrary fixed $k$, we may apply the same method to
$\partial \psi \slash \partial x$, proving that $\psi(t,\cdot,\cdot) \in H^1(\Omega)$.

Further, as $\psi$ evidently satisfies the specified boundary conditions, it remains to prove \eqref{psi}, which involves calculating  $\Delta \psi$ in  the distributional sense. Arguing as above, we find out that,
\begin{equation}
\label{Ldistribpsi}
-\Delta \psi(t,x,y) = \int_{\mathbb R} (\mu_{n_0}+k^2) a(k) e^{-i(\mu_{n_0}+k^2)t} f(k,x,y) dk,
\end{equation}
with the integrand lying in $L^2(\Omega)$. Since the r.h.s. of \eqref{Ldistribpsi} coincides with
$-i\partial\psi(t,x,y)/\partial t$ (the derivative is computed for the topology of the norm in $L^2(\Omega)$), we end up getting \eqref{psi}.
\begin{flushright} $\square$ \end{flushright}

Having established the main properties of the functions $\psi$ defined by \eqref{time1} and $\psi^{\pm}$ defined by \eqref{time2}--\eqref{time5} in Proposition \ref{pr-2}, we are in position to prove the main result of this section.

\begin{theorem}
\label{TLtheorem}
Let $a\in C^1_0({\mathbb R})$ obey \eqref{time0a}-\eqref{time0b}, let $\psi$ be the same as in \eqref{time1} and let $\psi^{\pm}$ be defined by \eqref{time2}--\eqref{time5}.
Then we have
\begin{equation}
\label{TLminus}
\lim_{t\to \pm \infty} \| \psi(t,\cdot,\cdot)-\psi^{\pm}(t,\cdot,\cdot) \|_{L^2(\Omega)} = 0.
\end{equation}
\end{theorem}
{\bf Proof.} We first examine the case of $t>0$. In light of \eqref{fleft}-\eqref{fright}, \eqref{time1}--\eqref{time5} and \eqref{psi2}-\eqref{psi1}, the function $\psi$ decomposes for any $x \in \mathbb{R}$ as
\begin{eqnarray}
\psi(\cdot,x,\cdot) & = & \theta(-x) \left( \psi^{-}(\cdot,x,\cdot) + \psi_{DN}(\cdot,x,\cdot) + \tilde{\psi}_{DN}(\cdot,x,\cdot) \right) \nonumber \\
& + &  \theta(x) \left( \psi_{ND}(\cdot,x,\cdot) + \tilde{\psi}_{ND}(\cdot,x,\cdot) \right) , \label{dec}
\end{eqnarray}
where $\theta$ stands for the usual Heaviside function, and $\psi^{-}$, $\psi_{DN}$, $\psi_{ND}$, $\tilde{\psi}_{DN}$ and $\tilde{\psi}_{ND}$ are respectively defined by \eqref{time2}, \eqref{time4}, \eqref{time5}, \eqref{psi1} and \eqref{psi2}.

Let us now establish that the transversal modes associated with $n \geq n_1+1$, appearing in the decomposition of $\psi(t,\cdot,\cdot)$, vanish as $t \to \pm \infty$. In view of \eqref{time1}, \eqref{time4}-\eqref{time5} and \eqref{dec}, it is enough to examine the functions $\tilde{\psi}_{DN}$ and $\tilde{\psi}_{ND}$. Applying Fubini Theorem, we obtain that
\begin{equation}
\label{Fnorm}
\|\tilde{\psi}_{ND}(t,\cdot,\cdot)\|_{L^2(\Omega)}^2 = \int_{\mathbb R_+} \sum_{n=n_1+1}^\infty \left| \int_{\mathbb R} a(k)  e^{-i(\mu_{n_0} + k^2)t} t_n(k) e^{-\kappa_n(k) x} dk \right|^2 dx,
\end{equation}
since $\{ \chi^{(+)}_n,\ n \geq 1 \}$ is an orthornomal basis of $L^2(0,d)$. Further we get
\begin{eqnarray*}
& & \int_{\mathbb R} a(k) e^{-i(\mu_{n_0} + k^2)t} t_n(k) e^{-\kappa_n(k) x} dk \\
& = & \frac{1}{2it} \int_{\mathbb R} e^{-i(\mu_{n_0} + k^2)t} \frac{\partial}{\partial k} \left( k^{-1} a(k) t_n(k) e^{-\kappa_n(k) x} \right) dk,\ x \in \mathbb{R}_+,\ t>0,\
\end{eqnarray*}
by integrating by parts.
Taking into account that $a \in C^1(\mathbb{R})$ is supported in $[\alpha, \beta]$, with $\alpha>0$, that
$t_n(k)$ satisfies the estimate (\ref{coeff_estim}), that
$$
\frac{d t_n}{d k}(k) = \frac{d}{d k}(\chi^{+}_n,\varphi(k))_{L^2(0,d)} = \left( \chi^{+}_n,\frac{\partial\varphi}{\partial k}(k) \right)_{L^2(0,d)},
$$
where $k \mapsto \frac{\partial \varphi}{\partial k}(k) \in C^0(\mathbb{R}_+^*,\cH^{\frac{1}{2}}$) from Theorem \ref{dphidk}, and that $\kappa_n(k) \geq \kappa_n(\beta)$ for all $k \in [\alpha,\beta]$, we find out two positive constants $d_1$ and $d_2$,
both of them being independent of $n$, such that the estimate
$$
\left| \int_{\mathbb R} a(k) e^{-i(\mu_{n_0} + k^2)t} t_n(k) e^{-\kappa_n(k) x} dk \right| \leq \frac{d_1+d_2 x}{2|t|\sqrt{n}}e^{-\kappa_n(\beta) x},
$$
holds uniformly in $t \in \mathbb{R}^*$ and $x \in \mathbb{R}_+^*$.
From this and (\ref{Fnorm}) then follows that
\begin{equation}
\label{Fnorm_estim2}
\| \tilde{\psi}_{ND}(t,\cdot,\cdot)\|_{L^2(\Omega)}^2 \leq \frac{c}{|t|},\ t \in {\mathbb R}^*,
\end{equation}
for some constant $c>0$, which does not depend on $t$. Arguing in the same way we get in addition that
\begin{equation}
\label{Fnorm_estim1}
\Vert \tilde{\psi}_{DN}(t,\cdot,\cdot)\|_{L^2(\Omega)}^2 \leq \frac{c}{|t|},\ t \in {\mathbb R}^*.
\end{equation}

The next step of the proof involves estimating $\| \psi^-(t,\cdot,\cdot) - \psi_{ND}(t,\cdot,\cdot) \|_{L^2({\mathbb R}_+^* \times (0,d))}$ for $t<0$. Recalling \eqref{time2} and
\eqref{time5}, we get
\begin{eqnarray*}
& & \psi^{-}(t,x,y)-\psi_{ND}(t,x,y) \\
& = & \int_{\mathbb R} a(k) e^{-i(\mu_{n_0}+k^2)t} \sum_{l=1}^{n_1} e^{ik_l x} \left(\delta_{l n_0} \chi^{-}_l(y) - t_l(k)\chi^{+}_l(y)\right) dk  \\
& = & -\sum_{l=1}^{n_1} \int_{\mathbb R} e^{i\left( k_l x - (\mu_{n_0}+k^2)t \right)}\frac{\partial q_l}{\partial k}(k,t,x,y) dk,
\end{eqnarray*}
upon integrating by parts, where
$$ q_l(k,t,x,y)=  \frac{a(k)}{i(-2kt+k k_l^{-1} x)}(\delta_{l n_0} \chi^{-}_l(y)
- t_l(k)\chi^{+}_l(y) ).$$
From this and (\ref{kl}) then follows that
$$ | \psi^{-}(t,x,y)-\psi_{ND}(t,x,y) | \leq \frac{e_1}{x-2k_{n_1}(\alpha)t} + \frac{e_2 x - e_3 t}{(x-2k_{n_1}(\alpha)t)^2}, $$
for $t <0$, $x >0$ and $y \in (0,d)$,
where $e_1$, $e_2$ and $e_3$ are three positive constants which are independent of $t$, $x$ and $y$. As a consequence we may find $c>0$, such that for all $t<0$, we have
\begin{equation}
\label{Fnorm_estim3}
\| \psi^-(t,\cdot,\cdot) - \psi_{ND}(t,\cdot,\cdot) \|_{L^2({\mathbb R}_+^*\times (0,d))}^2 \leq \frac{c}{|t|}.
\end{equation}
Analogously, recalling from Theorem \ref{dphidk} that  $k \mapsto r_l(k)$ and $k \mapsto t_l(k)$ are smooth for all $l=0,\ldots,n_1$, we get for every $t<0$ that
\begin{equation}
\label{Fnorm_estim4}
\| \psi_{DN}(t,\cdot,\cdot) \|_{L^2({\mathbb R}_-^* \times (0,d))}^2 \leq  \frac{c}{|t|}.
\end{equation}
Now, since
\begin{eqnarray*}
& & \| \psi(t,\cdot,\cdot)-\psi^{-}(t,\cdot,\cdot) \|_{L^2(\Omega)}^2 \\
& \leq &
\| \psi^{-}(t,\cdot,\cdot)-\psi_{ND}(t,\cdot,\cdot) \|_{L^2({\mathbb R}_+^* \times (0,d))}^2 +
\| \psi_{DN}(t,\cdot,\cdot)\|_{L^2({\mathbb R}_-^* \times (0,d))}^2  \\
& + &  \|\tilde{\psi}_{DN}(t,\cdot,\cdot)\|_{L^2(\Omega)}^2+  \| \tilde{\psi}_{ND}(t,\cdot,\cdot)\|_{L^2(\Omega)}^2,
\end{eqnarray*}
from \eqref{dec}, we obtain the desired result for $t \to -\infty$ by putting \eqref{Fnorm_estim2}--\eqref{Fnorm_estim4} together.

Similarly, arguing as above for $t>0$, by means of the decomposition
$$
\psi(\cdot,x,\cdot)=\psi^{+}(\cdot,x,\cdot) + \theta(x) \tilde{\psi}_{ND}(\cdot,x,\cdot) + \theta(-x) \left( \psi^{-}(\cdot,x,\cdot) + \tilde{\psi}_{DN}(\cdot,x,\cdot) \right),
$$
we obtain the desired result for  $t \to +\infty$.
\begin{flushright} $\square$ \end{flushright}

As seen in the proof of (\ref{TLminus}), $\| \psi(t,\cdot,\cdot)-\psi^{\pm}(t,\cdot,\cdot) \|_{L^2(\Omega)}$ scales like ${\mathcal O}(t^{-1/2})$. Actually, this is due to the fact that $a$ is $C^1$, as a smoother $a$ would allow for several integrations by parts, and consequently for a faster time decay.

Further, assumption \eqref{time0a} guarantees that the quasi-momenta involved in the framework of in Theorem \ref{TLtheorem} remain separated from the discrete threshold momentum values of $M$. This is not a very strong
restriction when studying the scattering of particles with momentum close to given value of $k>0$, which is sufficient for the definition of
reflection and transmission coefficients, as the skipped threshold points form a zero measure set where the reflection and transmission coefficients need not necessarily be defined.
However, their behavior near the thresholds, which we do not attempt to study in this paper, is certainly a point of interest.

\section{Numerical results}
For the illustration, we present here also numerically computed transition probabilities. Recall that reflection and transmission probability densities from the initial state of transversal mode
$n_0$ and longitudinal momentum $k$ to the final transversal mode $m$ are
$$
PRm=\frac{k_m(k)}{k} |r_m(k)|^2 \quad, \quad PTm=\frac{k_m(k)}{k} |t_m(k)|^2 \quad.
$$
The matching conditions were cut to the finite number of transversal modes, projected on a suitable finite-dimensional basis, and the resulting system of linear equations for the reflection and transmission coefficients solved numerically. No attempt to prove the convergence of this procedure was done, the precision was tested only numerically.

\begin{figure}[h]
\begin{center}
\includegraphics[width=0.8\textwidth,height=0.30\textheight]{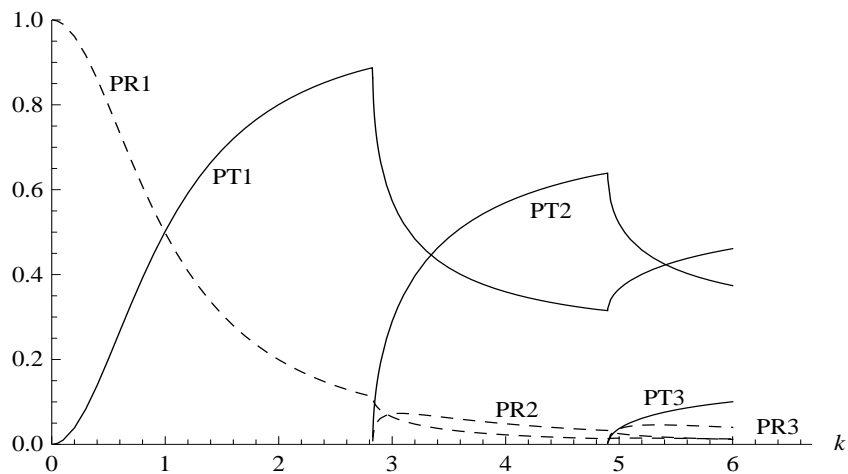}
\caption{Reflection (dashed) and transmission probabilities $PRm$ and $PTn$ from the initial transversal state $n_0=1$ and longitudinal momentum $k$ (in units of $\frac{\pi}{2d}$) to the transversal states $m$ and $n$ respectively. }
\end{center}
\end{figure}

\begin{figure}[h]
\begin{center}
\includegraphics[width=0.8\textwidth,height=0.30\textheight]{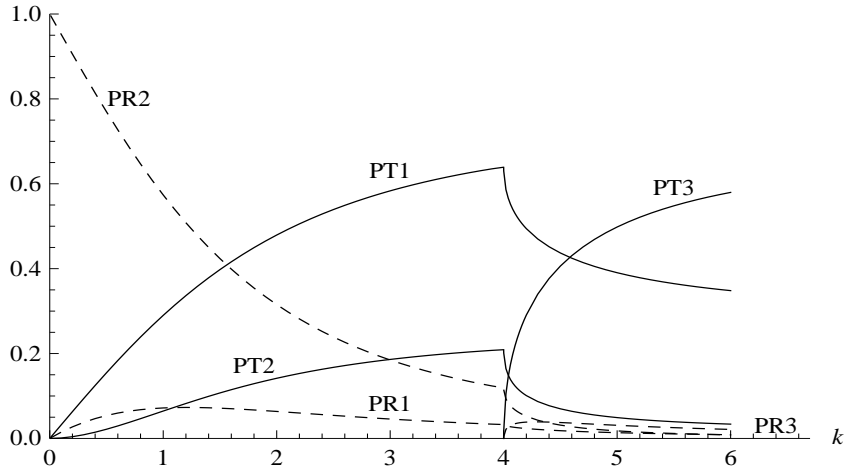}
\caption{Reflection (dashed) and transmission probabilities $PRm$ and $PTn$ from the initial transversal state $n_0=2$ and longitudinal momentum $k$ (in units of $\frac{\pi}{2d}$) to the transversal states $m$ and $n$ respectively. }
\end{center}
\end{figure}

The computed probability densities for $n_0=1,2$ are shown in Figures 2 and 3. They seem to be continuous but not smooth at the thresholds where excitations of the further transversal modes in the final state are opened. The total reflection at
$k\to 0$ is seen in accordance with the Dirichlet decoupling at $d\to 0$ proved by Borisov and Cardone \cite{Borisov} (notice, that the dimesionless value of  $kd$ is a relevant parameter).

%%%%%%%%%%%%%%%%%%%%%%%%%%%%%%%%%%Appendix%%%%%%%%%%%%%%%%
\section{Appendix}
\subsection{Proof of lemma \ref{lm-D2}}
The proof is by means of the two following technical lemmas.
With the notation (\ref{projectors}), we shall also write $P^\pm_N=\sum_{n=1}^N p^\pm_n$.

\subsubsection{Preliminary results}

\begin{lemma}
\label{lm-pre}
Let $N \in \mathbb{N}^*$. Then there exists $\epsilon_N \in (0,1)$ such that we have
$$ \min \left( \|P_N^+ \varphi\|_{L^2(0,d)} , \|P_N^- \varphi\|_{L^2(0,d)} \right) \leq \varepsilon_N \|\varphi\|_{L^2(0,d)},\ \varphi \in \mathcal{P}_N.$$
\end{lemma}
{\bf Proof.} Let us first establish that $\{ \chi_n^\pm,\ n=1,\ldots, N \}$
is linearly independent. We consider $2N$ complex numbers $\alpha_n^\pm$ for $n=1,\ldots, N$ such that
$\sum_{n=1}^N (\alpha_n^+ \chi^{+}_n + \alpha_n^- \chi^{-}_n ) = 0$.
Differentiating $2r$-times, $r \in \mathbb{N}$, we get that
$$ \sum_{n=1}^N \mu_n^r (\alpha_n^+ \chi^{+}_n + \alpha_n^- \chi^{-}_n ) = 0. $$
Evaluating this identity at $x=0$, we obtain $\sum_{n=1}^N \mu_n^r \alpha_n^+ = 0$.
The determinant of the system formed by the above equations with $r=0,\dots,N-1$, is Vandermonde and equals
$\prod_{1 \leq r < s \leq N} (\mu_s-\mu_r) > 0$.
Therefore $\alpha_n^+=0$ for all $n=1,\ldots,N$. Arguing as before with $x=d$ we obtain that $\alpha_n^-=0$ for $n=1,\ldots,N$, showing that the vectors $\chi_n^\pm$, for $n=1,\ldots, N$, are linearly independent.

Assuming that the statement of the Lemma does not hold, there would be
a sequence $\{ \varphi_n,\ n \in \mathbb{N}^* \} \in ({\mathcal P}_N)^{{\mathbb N}^*}$ such that
$$ \|\varphi_n\|_{L^2(0,d)}=1\ \mbox{and}\ \min(\|P_N^+ \varphi_n\|_{L^2(0,d)} , \|P_N^- \varphi_n\|_{L^2(0,d)} )  > 1-\frac{1}{n},$$
for every $n \in {\mathbb N}^*$. As the unit sphere in the finite dimensional linear space ${\mathcal P}_N$ is compact, we may find a subsequence that converges to a limit
$\varphi_0 \in {\mathcal P}_N$ obeying
$$ \|\varphi_0\|_{L^2(0,d)}=\|P_N^+ \varphi_0\|_{L^2(0,d)}=\|P_N^- \varphi_0\|_{L^2(0,d)}=1.$$
Thus we have $P_N^+ \varphi_0=\varphi_0=P_N^- \varphi_0$. Since $\{ \chi_n^\pm,\ n=1,\ldots, N \}$ is linearly
independent, then $\varphi_0=0$, contradicting the fact that $\|\varphi_0\|_{L^2(0,d)}=1$. As a consequence the statement is true for every $\varphi \in \mathcal{P}_N$ such that $\|\varphi\|_{L^2(0,d)}=1$.
Therefore it holds for any $\phi \in \mathcal{P}_N$ since it is valid for $\phi=0$ and for
$\phi \slash \|\phi\|_{L^2(0,d)}$ when $\phi \neq 0$.
\begin{flushright} $\square$ \end{flushright}

\begin{lemma}
\label{maxdown}
Let $N \in \mathbb{N}^*$. Then there exists $\eta_N \in (0,1)$ such that the estimate
$$
\max \left(  \|(I-P_N^+ ) \phi\|_{L^2(0,d)} , \|(I-P_N^-)\phi\|_{L^2(0,d)} \right) \geq \eta_N \|\phi\|_{L^2(0,d)},
$$
holds for all $\phi \in L^2(0,d)$.
\end{lemma}
{\bf Proof.} The proof boils down to the fact that
\begin{eqnarray*}
\| (I-P_N^\pm) \phi\|_{L^2(0,d)}^2 & =  & \| \phi\|_{L^2(0,d)}^2 - \| P_N^\pm \phi\|_{L^2(0,d)}^2 \\
&  = & \| \phi\|_{L^2(0,d)}^2 - \| P_N^\pm P_N \phi\|_{L^2(0,d)}^2,
\end{eqnarray*}
and $\| \phi\|_{L^2(0,d)}^2 = \| (I-P_N)  \phi\|_{L^2(0,d)}^2 + \| P_N \phi\|_{L^2(0,d)}^2$.
Indeed, this entails
\begin{eqnarray*}
& & \max \left( \|(I-P_N^+) \phi\|_{L^2(0,d)}^2 , \|(I-P_N^-)\phi\|_{L^2(0,d)}^2 \right) \\
&= & \|(I-P_N)\phi\|_{L^2(0,d)}^2 + \|P_N \phi\|_{L^2(0,d)}^2 \\
& & - \min \left( \|P_N^+ P_N \phi\|_{L^2(0,d)}^2 , \|P_N^- P_N \phi\|_{L^2(0,d)}^2 \right),
\end{eqnarray*}
which, together with Lemma \ref{lm-pre}, yields
\begin{eqnarray*}
& & \max \left( \|(I-P_N^+) \phi\|_{L^2(0,d)}^2 , \|(I-P_N^-)\phi\|_{L^2(0,d)}^2 \right) \\
& \geq & \|(I-P_N)\phi\|_{L^2(0,d)}^2 + (1-\epsilon_N^2)  \| P_N \phi\|_{L^2(0,d)}^2.
\end{eqnarray*}
Since the right hand side of the above estimate is lower bounded by $(1-\epsilon_N^2)  \| \phi\|_{L^2(0,d)}^2$, we obtain the desired result.
\begin{flushright} $\square$ \end{flushright}

% %%%%%%%%%%%%%%%%%%%%%%%%%%%%%%%%%%%%%%%%%%%%%%%%%%%%%%%%%%%%%%%%%%%%%%%%%%%%%%%%%
\subsubsection{Completion of the proof}
\label{sec-prlmD2}
For all $\phi \in \cH^{\frac{1}{2}}$, we have
$$ ( D_2 \phi , \phi )_{-\frac{1}{2} , \frac{1}{2}}=\sum_{n=n_1+1}^\infty \kappa_n \left( (p_n^- \phi,\phi )_{L^2(0,d)} + (p_n^+ \phi , \phi)_{L^2(0,d)} \right), $$
from (\ref{Ddef}), so we get
\begin{eqnarray}
( D_2 \phi , \phi )_{-\frac{1}{2} , \frac{1}{2}} & \geq &
\kappa_{n_1+1} \left( \|(I-P_{n_1}^-)\phi\|_{L^2(0,d)}^2+\|(I-P_{n_1}^+)\phi\|_{L^2(0,d)}^2 \right) \nonumber \\
& \geq &  \kappa_{n_1+1} \eta_{n_1}^2 \|\phi\|_{L^2(0,d)}^2, \label{lowbound}
\end{eqnarray}
by Lemma \ref{maxdown}, showing that $D_2$ is injective.

To prove that $D_2$ is a surjection, we first establish the two following lemmas.
\begin{lemma}
\label{Fproperties}
For each $\psi \in \cH^{-\frac{1}{2}}$, the functional
$$ F_{\psi}(\phi)=( D_2 \phi , \phi )_{-\frac{1}{2} , \frac{1}{2} } -  2 \Re ( \psi, \phi)_{-\frac{1}{2} , \frac{1}{2} }
$$
is real-valued, strictly convex, continuous and coercive on $\cH^{\frac{1}{2}}$.
\end{lemma}
{\bf Proof.} First, it is apparent from the definition of $F_\psi$ that $F_\psi(\phi) \in \mathbb{R}$ for all $\phi \in \cH^{\frac{1}{2}}$.

Next, we get through direct calculation that
\begin{eqnarray*}
& & t F_\psi(\phi_1) + (1-t) F_\psi(\phi_2) - F_\psi(t\phi_1 + (1-t)\phi_2) \\
&=&  t(1-t)(D_2 ( \phi_1-\phi_2) , \phi_1-\phi_2 )_{-\frac{1}{2} , \frac{1}{2} },
\end{eqnarray*}
for each $t \in [0, 1]$ and all $\phi_1, \phi_2 \in \cH^{\frac{1}{2}}$. This and (\ref{lowbound}) yield that $F_\psi$ is strictly convex.

Further, $\| D_2 \delta \|_{-\frac{1}{2}}$ being majorized, up to some multiplicative positive constant, by $\| \delta \|_{\frac{1}{2}}$, uniformly in $\delta=\sum_{n=1}^{+\infty} \delta_n^\pm \chi_n^\pm \in \cH^{\frac{1}{2}}$,  as can be seen from the following computation arising from
(\ref{kn}),
\begin{eqnarray*}
\| D_2 \delta \|_{-\frac{1}{2}}^2 & = & \sum_{n=n_1+1}^{+\infty} \frac{\kappa_n^2}{n} \left( | \delta_n^+ |^2 + | \delta_n^- |^2 \right) \\
& \leq & \frac{\pi^2}{d^2} \sum_{n=n_1+1}^{+\infty} \frac{n^2-n_0^2}{n} \left( | \delta_n^+  |^2 + | \delta_n^- |^2 \right) \\
& \leq & \frac{\pi^2}{d^2} \sum_{n=n_1+1}^{+\infty} n \left( | \delta_n^+|^2 + | \delta_n^- |^2 \right) \leq \frac{\pi^2}{d^2}\| \delta \|_{\frac{1}{2}}^2,
\end{eqnarray*}
the continuity of $F_\psi$ follows readily from the basic estimate
\begin{eqnarray*}
& & | F_\psi(\phi+\delta)-F_\psi(\phi) |  \\
& \leq &  2 \left| \Re (D_2 \phi, \delta)_{-\frac{1}{2} , \frac{1}{2}} - \Re (\psi , \delta)_{-\frac{1}{2} , \frac{1}{2} } \right| + (D_2 \delta,\delta)_{-\frac{1}{2} ,  \frac{1}{2}} \\
& \leq &
2 \left( \|D_2\phi\|_{-\frac{1}{2}} + \|\psi\|_{-\frac{1}{2}}  + \| D_2 \delta \|_{-\frac{1}{2}} \right) \|\delta\|_{\frac{1}{2}},
\end{eqnarray*}
which holds true for any $\phi \in \cH^{\frac{1}{2}}$.

Let us now prove that $F_\psi$ is coercive. We use (\ref{kn}), getting $\kappa_n \geq \frac{\kappa_{n_1+1}}{n_1+1} n$ for every $n \geq n_1+1$, hence
\begin{equation}
\label{D2increase}
(D_2 \phi, \phi)_{-\frac{1}{2},\frac{1}{2}} \geq \frac{\kappa_{n_1+1}}{n_1+1} \sum_{n=n_1+1}^\infty n(|a_n^+|^2 + |a_n^-|^2),
\end{equation}
for all $\phi = \sum _{n=1}^\infty a_n^\pm \chi_n^{\pm} \in \cH^{\frac{1}{2}}$.
Further, since
$$\sum_{n=n_1+1}^{+\infty} n ( | a_n^+ |^2 + | a_n^- |^2 ) \geq (n_1+1) \sum_{n=n_1+1}^{+\infty} ( | a_n^+ |^2 + | a_n^- |^2 ), $$
we derive from Lemma \ref{maxdown} that
\begin{eqnarray*}
\sum_{n=n_1+1}^{+\infty} n ( | a_n^+ |^2 + | a_n^- |^2 ) & \geq & \frac{n_1+1}{2} \eta_{n_1}^2 \sum_{n=1}^{+\infty} ( | a_n^+ |^2 + | a_n^- |^2 ) \\
& \geq & \frac{n_1+1}{2 n_1} \eta_{n_1}^2 \sum_{n=1}^{n_1} n ( | a_n^+ |^2 + | a_n^- |^2 ).
\end{eqnarray*}
Bearing in mind that $\eta_{n_1} \in (0,1)$, this entails that
$$ \sum_{n=n_1+1}^{+\infty} n ( | a_n^+ |^2 + | a_n^- |^2 ) \geq \frac{n_1+1}{3 n_1+1} \eta_{n_1}^2 \| \phi \|_{\frac{1}{2}}^2, $$
for all $\phi \in \cH^{\frac{1}{2}}$.
From this and (\ref{D2increase}) then follows that
$$
(D_2 \phi, \phi)_{-\frac{1}{2},\frac{1}{2}} \geq \frac{\kappa_{n_1 + 1}}{3 n_1+1} \eta_{n_1}^2 \|\phi\|_{\frac{1}{2}}^2,\ \phi \in \cH^{\frac{1}{2}},
$$
showing that $F_\psi$ is coercive.
\begin{flushright} $\square$ \end{flushright}

\begin{lemma}
\label{D2eqlemma}
Let $\psi$ and $F_\psi$ be the same as in Lemma \ref{Fproperties}.
Assume that $F_\psi$ has its minimum at $\phi \in \cH^{\frac{1}{2}}$. Then we have
\begin{equation}
\label{D2equation}
D_2 \phi = \psi\ in\ \cH^{-\frac{1}{2}}.
\end{equation}
\end{lemma}
{\bf Proof.} Fix $\delta\in \cH^{\frac{1}{2}}$. The function $\mathbb{R} \ni t \mapsto F_\psi(\phi+t\delta)$
has a minimum at $t=0$, by assumption, so its derivative is zero there. As
$$
F_\psi(\phi+t\delta) = F_\psi(\phi) + 2 t \Re (D_2 \phi - \psi , \delta )_{-\frac{1}{2},\frac{1}{2}}  + t^2 ( D_2 \delta , \delta )_{-\frac{1}{2} , \frac{1}{2}},
$$
this means that $\Re  ( D_2 \phi -\psi ,\delta )_{-\frac{1}{2} , \frac{1}{2}}  = 0$. Further, substituting $i \delta$ for $\delta$ in this equation, we find out that
$$ ( D_2 \phi - \psi ,\delta )_{-\frac{1}{2} , \frac{1}{2} } =0, $$
which entails (\ref{D2equation}) since $\delta$ is arbitrary in $\cH^{\frac{1}{2}}$.
\begin{flushright} $\square$ \end{flushright}

We are now in position to prove the first assertion in Lemma \ref{lm-D2}.
For $\psi$ in $\cH^{-\frac{1}{2}}$ fixed, we know from Lemma \ref{Fproperties} that $F_\psi$ admits a unique minimum $\phi \in \cH^{\frac{1}{2}}$ (see e.g. \cite{Jaffe}), which is solution
to (\ref{D2equation}), according to Lemma \ref{D2eqlemma}. This establishes that $D_2$ is surjective, and thus one-to-one
from $\cH^{\frac{1}{2}}$ onto $\cH^{-\frac{1}{2}}$, since it is already known to be injective. Moreover, the boundedness
of $D_2^{-1}$ follows from the inverse mapping theorem (e.g. \cite{RS4}[Theorem III.11]).

Last, we prove that the operator $(D_2^{-1})_{| L^2(0,d)}$ is  symmetric. We pick $f, g \in \cH^{-\frac{1}{2}}$ and recall from (\ref{Ddef}) that $(D_2 u , v)_{-\frac{1}{2} , \frac{1}{2}} = \overline{( D_2 v , u)}_{-\frac{1}{2},\frac{1}{2}}$ for every $u, v \in \cH^{\frac{1}{2}}$. Taking $u=D_2^{-1} f$ and $v=D_2^{-1} g$ in this equation we get that
\begin{equation}
\label{relsym}
( f , D_2^{-1} g )_{-\frac{1}{2},\frac{1}{2}}=\overline{( g , D_2^{-1} f)}_{ -\frac{1}{2}, \frac{1}{2} }.
\end{equation}
In the peculiar case where $f, g \in L^2(0,d)$, we have $( f , D_2^{-1} g )_{-\frac{1}{2},\frac{1}{2}}=( f , D_2^{-1} g )_{L^2(0,d)}$
and $( g , D_2^{-1} f)_{-\frac{1}{2} , \frac{1}{2} }=( g , D_2^{-1} f )_{L^2(0,d)}$, hence
$$ ( f , D_2^{-1} g )_{L^2(0,d)} = ( D_2^{-1} f, g )_{L^2(0,d)}, $$
from (\ref{relsym}), which is the required symmetry relation.

% %%%%%%%%%%%%%%%%%%%%%%%%%%%%%%%%%%%%%%%%%%%%%%%%%%%%%%%%%%%%%%%%%%%%%%%%%%%%%%%%%
\subsection{Limits in matching conditions}
\label{sec-scndmatching}
This appendix is devoted to proving that the limit \eqref{match0} coincides with the right hand side of \eqref{match1}.
%We shall systematically omit the argument $k$ in the function $f$ in this appendix.
To do that we start by establishing that
\begin{enumerate}[(i)]
\item $\frac{\partial \varphi}{\partial x}(\pm\varepsilon,\cdot)$ goes to $\frac{\partial \varphi}{\partial x}(0,\cdot)$ in $L^2(0,d)$ as $\varepsilon \downarrow 0$;
\item $f(k,\pm\varepsilon,\cdot)$ tends to $f(k,0^\pm,\cdot)$ in $L^2(0,d)$ as $\varepsilon \downarrow 0$.
\end{enumerate}
First, we notice for every $y \in (0,d)$ and $\varepsilon>0$ that
$$
\left| \frac{\partial\varphi}{\partial x}(\pm\varepsilon,y) - \frac{\partial\varphi}{\partial x}(0,y) \right| \leq
\left\|\frac{\partial^2 \varphi}{\partial x^2}\right\|_{L^{\infty}(\Omega)} \varepsilon,
$$
so we get
$$
\left\| \frac{\partial\varphi}{\partial x}(\pm\varepsilon,\cdot) - \frac{\partial\varphi}{\partial x}(0,\cdot) \right\|_{L^2(0,d)} \leq
\left\|\frac{\partial^2 \varphi}{\partial x^2}\right\|_{L^{\infty}(\Omega)} \sqrt{d} \varepsilon ,
$$
which entails (i).

For the second statement (ii), we shall only prove that $\lim_{\varepsilon \downarrow 0} f(k,-\varepsilon,\cdot)=f(k,0^-,\cdot)$ in $L^2(0,d)$, the case of the limit from the right being fully analogical.
From the definition of the trace $f(k,0^-,\cdot)$, we may find $L>0$,  $\varepsilon_0 >0$ and a sequence of functions $(f_n)_{n \geq 1}$ in $C^\infty(\Omega)$ satisfying
\begin{equation}
\label{limaco1}
\lim_{n \to +\infty} \| f_n - f(k,\cdot,\cdot) \|_{H^1((-L-\varepsilon_0,0)\times (0,d))} =0,
\end{equation}
such that $f_n(0,\cdot)$ tends to $f(k,0^-,\cdot)$ in $\cH^{\frac{1}{2}}$ as $n$ goes to infinity. Moreover we have
$$
\| f_n(0,\cdot) - f(k,0^-,\cdot)\|_{\frac{1}{2}} \leq c \|f_n - f(k,\cdot,\cdot)\|_{H^1((-L,0)\times (0,d))},
$$
for some positive constant $c$ which is independent of $f$ and $f_n$. Further, since $\|f(k,-\varepsilon,\cdot) - f(k,0^-,\cdot)\|_{\frac{1}{2}}$ is majorized by the sum
$$ \|f(k,-\varepsilon,\cdot) - f_n(-\varepsilon,\cdot)\|_{\frac{1}{2}} +
\|f_n(-\varepsilon,\cdot) - f_n(0,\cdot)\|_{\frac{1}{2}} +
\|f_n(0,\cdot) - f(k,0^-,\cdot)\|_{\frac{1}{2}}, $$
for every $\varepsilon \in (0,\varepsilon_0)$, we get that
\begin{eqnarray}
& & c^{-1} \|f(k,-\varepsilon,\cdot) - f(k,0^-,\cdot)\|_{\frac{1}{2}} \nonumber \\
& \leq & \|f(k,\cdot,\cdot)-f_n\|_{H^1((-L-\varepsilon, -\varepsilon)\times (0,d))} +
\|f_n(\cdot -\varepsilon,\cdot)-f_n(\cdot,\cdot)\|_{H^1((-L,0)\times (0,d))} \nonumber \\
& + & \|f_n-f(k,\cdot,\cdot)\|_{H^1((-L,0)\times (0,d))}. \label{limaco2}
\end{eqnarray}
In light of \eqref{limaco1}, the first and third terms in the right hand side of \eqref{limaco2} can de made arbitrarily small upon chosing $n$
sufficiently large. For such an $n$, using the uniform continuity of the function $f_n$ or applying \cite{Zuily}[Chapter 1, Lemma 3.5], it is possible
to make the second term arbitrarily small by taking $\varepsilon>0$ small enough. As a consequence we have
$\lim_{\varepsilon \downarrow 0} f(k,-\varepsilon,\cdot)=f(k,0^-,\cdot)$ in $\cH^{\frac{1}{2}}$, and thus in $L^2(0,d)$, which is the statement of (ii).

Putting (i) and (ii) together, we obtain that
\begin{eqnarray}
& & \lim_{\varepsilon\downarrow 0} \left( \int_0^d \bar{f}(k,-\varepsilon,y) \frac{\partial \varphi}{\partial x}(-\varepsilon,y)dy - \int_0^d \bar{f}(k,\varepsilon,y) \frac{\partial \varphi}{\partial x}(\varepsilon,y)dy \right) \nonumber \\
& = &  \left( f(k,0^-,\cdot)-f(k,0^+,\cdot) , \frac{\partial \varphi}{\partial x}(0,\cdot)\right)_{L^2(0,d)}. \label{limaco3}
\end{eqnarray}

The rest of the proof is to compute $\int_0^d \frac{\partial \bar{f}}{\partial x}(k,\pm \varepsilon,\cdot) \varphi(\pm \varepsilon,y) dy$ as $\varepsilon \downarrow 0$. We shall actually restrict ourselves to calculating the limit from the left, the other case being fully analogical. To do that we start by decomposing $\varphi(-\varepsilon,\cdot) \in \cH^{\frac{1}{2}}_-$, as
$$
\varphi(-\varepsilon,\cdot) = \sum_{n=1}^\infty b_n(-\varepsilon)\chi^{-}_n,\ b_n(-\varepsilon)= \int_0^d \chi^{-}_n(y) \varphi(-\varepsilon,y) dy,\ n \in \mathbb{N}^*.
$$
and notice from (\ref{chidef}) upon integrating several times by parts, that
$$
|b_n(-\varepsilon)|=\left| \int_0^d \chi^{-}_n(y) \varphi(-\varepsilon,y) \, dy \right| \leq
 \frac{c}{\mu_n^2},\ c=\sqrt{2d} \left\|\frac{\partial^4\varphi}{\partial y^4}\right\|_{L^{\infty}(\Omega)}.
$$
Further, it follows from
(\ref{fleft}) and (\ref{F-trace}) that
\begin{eqnarray}
& &  \int_0^d \frac{\partial \bar{f}}{\partial x}(k,-\varepsilon,y) \varphi(-\varepsilon,y) dy -  ( f_x(k,0^-,\cdot) , \varphi(0,\cdot) )_{-\frac{1}{2}, \frac{1}{2}} \\
& = &  -ik(b_{n_0}(0)-b_{n_0}(-\varepsilon)e^{-ik\varepsilon}) +i\sum_{l=1}^{n_1}k_l \overline{r_l} (b_l(0)-b_l(-\varepsilon))e^{ik_l\varepsilon}) \nonumber \\
& -  & \sum_{n=n_1+1}^\infty \kappa_n \overline{r_n} (b_n(0)-b_n(-\varepsilon)e^{-k_n\varepsilon}), \label{limaco4}
\end{eqnarray}
from where we get
\begin{eqnarray}
& & \left|  \int_0^d \frac{\partial \bar{f}}{\partial x}(k,-\varepsilon,y) \varphi(-\varepsilon,y) dy -  ( f_x(k,0^-,\cdot) , \varphi(0,\cdot) )_{-\frac{1}{2}, \frac{1}{2}} \right| \\
& \leq &
\frac{2ck}{\mu_{n_0}^2} + 2c\sum_{l=1}^{n_1}\frac{k_l |r_l|}{\mu_l^2} + 2 c \sum_{n=n_1+1}^\infty \frac{\kappa_n |r_n| }{\mu_n^2}.
\end{eqnarray}
Bearing in mind that $\mu_n$ scales like $n^2$, and recalling \eqref{kn} and \eqref{S1_2}, we see that the series in the above expression is convergent. Thus we obtain
$$
\lim_{\varepsilon \downarrow 0} \int_0^d \frac{\partial \bar{f}}{\partial x}(k,-\varepsilon,y) \varphi(-\varepsilon,y) dy = ( f_x(k,0^-,\cdot) , \varphi(0,\cdot) )_{-\frac{1}{2}, \frac{1}{2}},
$$
by taking the limit $\varepsilon\to 0^+$ term by term in \eqref{limaco4} and using the continuity of $\varepsilon \mapsto b_n(-\varepsilon)$ at $0$.
Now the desired result follows from this and \eqref{limaco3}.
\\
%\vspace{1cm}
\\
{\bf Acknowledgement}. The work was partly supported by the Czech Science Foundation project 14-06818S and by the NPI ASCR institutional
support RVO 61389005. J.D. is indebted to CPT CNRS Marseille for the hospitality extended to him
during several visits.

\end{document}